\documentclass[aps,epl,twocolumn,superscriptaddress,groupedaddress,showpacs]{revtex4-1}
\usepackage{graphicx}
\usepackage{amsmath}
\usepackage{amsfonts}
\usepackage{color}

\begin{document}

\title{Field Induced Oscillation of Two Majorana Modes in a Quantum Ring}
\author{Yue Yu}
\affiliation{-2017, Department of Physics, Hong Kong University of Science and Technology, Hong Kong}
\affiliation{2017-, Stanford University, Stanford }
\author{Kwok Yip Szeto}
%\thanks{}
\affiliation{Department of Physics, Hong Kong University of Science and Technology, Hong Kong}
\date{\today}
\begin{abstract}

We calculate the topological boundary modes found for quantum wire for a quantum ring. For the symmetric ring we find analytical solutions for two quasi-particles identifiable as the Majorana states and for asymmetric ring, we have also find approximate solution. By turning on a small field at time zero, we find a field induced oscillation of the Majorana states that was initially localized. The range of validity of our solution, the relation between the period of oscillation and the coin parameter and field strength, and the encoding of a message of the on/off of the field using the Majorana states are discussed.   Our results suggest ways to reduce decoherence for small systems, thereby providing a good candidate for quantum memory and computing as the Majorana  qubit is topologically protected.

\end{abstract}
% insert suggested PACS numbers in braces on next line \pacs{}
%\keywords{}
\maketitle

\section{Introduction}
The recent advances in discrete time quantum walk \cite{A4,A5,B1,B2,B3} have created many interesting research endeavors in its application, such as in quantum computation \cite{A6,A7,A9,A10,A11}. 
However, finding a qubit that is free from decoherence is a major challenge. Recently, we see progress in this challenging problem 
with our increased understanding of topologically protected systems \citep{Nielsen, Nayak}. 
One of the simplest systems supporting these topological phases, which are identified as Majorana boundary modes \citep{Kitagawa, Hotat,B11,B12,B13,B14,B15,B16}, is the Kitaev model \citep{Kitaev} in the literature of discrete quantum walk\cite{B4,B5,B6,B7,B8,B9,B10}, for quantum wire. 
In this paper we use the exact solution for these modes in one dimension \citep{Hotat} to compute the response of the quantum walker to the stepwise increment of vector potential.  The general solutions of this problem involve numerical solutions of a set of coupled nonlinear equations, which permits analytical solution under certain limiting cases of long chain and small increment of vector potential. 
The limit of validity of the analytical solution turns out to be easily satisfied for most cases and thus provides a useful guideline for the experimentalists. 
Furthermore, we find interesting field induced oscillation that allows the preparation of the decoherence-free qubit. 
Here we investigate the interesting case of a ring, where the field induced oscillation can be controlled properly for application. 

For quantum walk in one dimension, we can find the low energy boundary modes which can be identified as symmetry protected Majorana bound states by setting parameters in the coin matrix properly \cite{Hotat}. With this solution, we study the interaction of the two Majorana bound states, using a coin matrix with a new phase $\alpha$ that corresponds to the addition of a vector potential. 
Firstly, the evolution of the wave function in one-dimensional discrete quantum walk is given by the unitary transformation $ U = S \otimes C$, which composes of the coin operator $C$ and a shifting operator 
$S = |L\rangle{\langle}L|\otimes\sum|n\rangle{\langle}n+1|+|R\rangle{\langle}R|\otimes\sum|n\rangle{\langle}n-1|$. 
Here, $|L,R\rangle$ are the basis of the coin space, while $|n\rangle$ are the basis of the position space. We focus on the coin operator in the following form 
\begin{equation}
C(\theta,\alpha)=\left( \begin{array}{cc}
\cos{\theta}e^{-i\alpha} & \sin{\theta}e^{-i\alpha} \\
-\sin{\theta}e^{i\alpha} & \cos{\theta}e^{i\alpha} \end{array} \right).\label{Eq:1}
\end{equation}
If the coin matrix is the same at every position in the system, we can express the energy eigenstates in the form a set of plane waves due to the translational symmetry. By substituting the eigenstate into the unitary operator, we find the dispersion relationship of E and k,
\begin{equation}
\begin{aligned}
\psi_{k}(n,t) = e^{-iEt+ikn}\left[\begin{array}{ccc}
a_k\\
b_k \end{array} \right], \ 
\cos(E)=\cos{\theta}\cos(q),\label{Eq:3}
\end{aligned}
\end{equation}
where $q=k-\alpha$. 
For a given energy level, there are two correspondent eigen-momentum, and a general wave function is the linear combination of the correspondent  eigenvector $[a_k\,b_k]^{T}$ \citep{Hotat2}, 
\begin{equation}
\left[\begin{array}{c}
a_k \\
b_k\end{array} \right]
= \frac{1}{\sqrt{2-2\cos\theta\cos(E+q)}}
\left[\begin{array}{c}
\sin\theta \\
(e^{-i(E+q)}-\cos\theta)\end{array} \right].\label{Eq:4}
\end{equation}
According to the discussion in \cite{Kitagawa} on an infinite wire, one can identify Majorana bound state, by setting topological boundary ($\theta>0$ for site $i\leq0$, $\theta<0$ for site $i<0$). 
We note that this observation on the infinite wire which is assumed to be homogeneous except at the boundary must be generalized as in real system there are impurities that can cause decoherence. 
Here, we analyze the eigenstate for quantum walk on a finite ring with two topological boundaries and hope that the finite size can mitigate the problem of decoherence.

\section{Eigenstate without $\alpha$ field}
\begin{figure}[h]
\centering
\includegraphics[width=6cm]{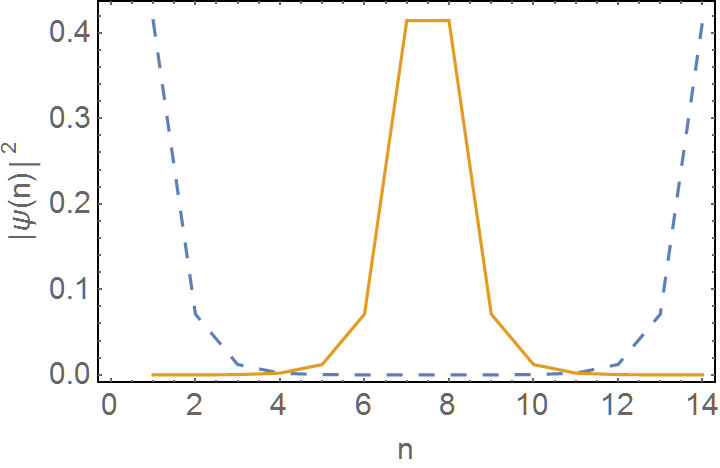}
\caption{The probability distribution $|\psi(n)|^2$ of the two degenerate $E=0$ Majoranas, on the symmetric ring $N_1=N_2=7$. $\theta$ is set to be $0.25\pi$. The two Majoranas are localized at the two topological boundaries $n=0$ (equivalently $n=N_1+N_2=14$ (blue dotted line) ), and $n=N_1=7$ (red line), respectively.}\label{F1}
\end{figure}

We begin with a finite ring with length $N_1+N_2$ without any field ($\alpha=0$). The sites on the ring are labelled by $n\in[0,N_1+N_2]$, with site $n=0$ identified with $n=N_1+N_2$. The coin parameter $\theta>0$ is positive for $n\in[0,N_{1}-1]$ and is ( $-\theta<0$ ) for $n\in[N_1,N_1+N_2-1]$. In this way, we have two topological boundaries at $n=0$ and $n=N_1$, and we anticipate two symmetry protected bound states \cite{Kitagawa, Kitaev, Hotat} in this ring. We now calculate the eigenstate and eigen-energy, by matching the boundary conditions. The general wave function is a linear combination of plane wave solutions, with amplitudes $u_1,u_2,u_3,u_4$, 
\begin{widetext}
\begin{equation}
\begin{split}
&\psi(n)=u_1e^{ikn}[a_k,b_k]^T+u_2e^{-ik(n-N_1)}[a_{-k},b_{-k}]^T,\,n\in[0,N_{1}-1]\\
&\psi(n)=u_3e^{ik(n-N_1)}[c_k,d_k]^T+u_4e^{-ik(n-N_1-N_2)}[c_{-k},d_{-k}]^T,\,n\in[N_1,N_1+N_2].
\end{split}
\end{equation}
\end{widetext}
Here $[a_{k},b_k]$ and $[c_k,d_k]$ are the eigenvectors for $\pm{\theta}$.  
The boundary conditions at $n=0, N_1 $ for left/right moving state provide four eigenstate equations
\begin{equation}
\begin{split}
C(\theta)\psi(N_1-1)_R=e^{-iE}\psi(N_1)_R\\
C(-\theta)\psi(N_1)_L=e^{-iE}\psi(N_1-1)_L\\
C(-\theta)\psi(N_1+N_2-1)_R=e^{-iE}\psi(0)_R\\
C(\theta)\psi(0)_L=e^{-iE}\psi(N_1+N_2-1)_L\\
\end{split}
\end{equation}
Here, R/L denotes the left moving ($a_k$ or $c_k$) and the right moving ($b_k$ or $d_k$) component. 
These boundary conditions provide four linear equations in matrix form $\bold A \vec u=0$ for the amplitudes $\vec u =[u_1,u_2,u_3,u_4]^T$, and the condition of nontrivial solution is that the determinant of the coefficient matrix $\bold A$ is zero. This zero determinant condition for matrix $\bold A$ leads to a quadratic equation of $e^{iE}$ which after simplification yield the exact solution for E
\begin{equation}
\begin{split}
\sin{E}=\pm\frac{(\mu^{-1}-\mu)(\mu^{N_1}-\mu^{N_2})\cos\theta}{2\sqrt{(1-\mu^{2N_1})(1-\mu^{2N_2})}}
\end{split}\label{E6}
\end{equation}
With $\mu=e^{ik}\in\mathbb{R}$ for the localization of the bound state. The solution for E and k can then be found by substituting the dispersion relationship and we get the exact solution for Majorana bound state solutions $E=0,\pi$ for the case of a symmetric ring $N_1=N_2$. Each Majorana state $E=0/\pi$ is doubly degenerate, and the two degenerate Majoranas are localized at the two topological boundaries $n=0$ and $n=N_1$(Fig.\ref{F1}). Note that these solutions for symmetric ring exist for any size of the system, including a small system. This suggests a possible way to reduce decoherence.

For asymmetric ring, we find that the energy will deviate from $0$ or $\pi$, but for large ring the deviation will be suppressed by the localization $\mu$ of the bound state. 
Let’s consider a large asymmetric ring so that ($\mu^{N_1},\mu^{N_2}\ll{1}$).
From the dispersion relation, we obtain the leading order approximation for $\mu$ 
\begin{equation}
\begin{split}
&\mu(E=\pm\epsilon)=\mu=\frac{\cos\theta}{1+\sin\theta}+O(\epsilon^2),\\
&\mu(E=\pi\pm\epsilon)=-\mu,
\end{split}
\end{equation}
After substituting this $\mu$ into Eq.\ref{E6}, we get the four bound state energies approximately  
\begin{equation}
\begin{split}
&E=\pm\epsilon_0, \pi\pm\epsilon_1,\,\,\,\mu=\frac{\cos\theta}{1+\sin\theta}\\
&\epsilon_0=\pm(\mu^{N_2}-\mu^{N_1})\sin\theta+o(\mu^{N_1}+\mu^{N_2}),\\
&\epsilon_1=\pm((-\mu)^{N_2}-(-\mu)^{N_1})\sin\theta+o(\mu^{N_1}+\mu^{N_2}).
\end{split}
\end{equation}
One can verify that $\epsilon_{0,1}\ll{1}$ for $\mu^{N_1},\mu^{N_2}\ll{1}$, so that the deviation from the solution to symmetric ring is small.

\section{Eigenstate with $\alpha$ field}
Here we want to manipulate the two bound states by an external perturbation of a constant uniform $\alpha$ field. 
We first consider another system of the one dimensional line with translational invariance and known eigenstate $\psi_{\alpha}$ for a given $\alpha$. There exists a simple relation between the eigenstate if $\alpha$ is changed to $\alpha'$ uniformly for every points in the system as $\psi_{\alpha'}(n,t)=\psi_{\alpha}(n,t)e^{i(\alpha'-\alpha)n }$. 
The relative phase between the two neighboring points will compensate the effect of $\alpha'-\alpha$ in the coin matrix, so that
\begin{equation}
\begin{aligned}
&(C_{\alpha'}\psi_{\alpha'}(n,t))_L=\psi_{\alpha'}(n-1,t+1)_L,
\\&(C_{\alpha'}\psi_{\alpha'}(n,t))_R=\psi_{\alpha'}(n+1,t+1)_R.
\end{aligned}\label{E9}
\end{equation}
This implies that the eigen-energy and the parameter $q$ are unchanged, but the quasi-momentum $k=q+\alpha$ is replaced by $k'=q+\alpha'$. 

Since a finite ring can be treated as an infinite line with periodic boundary conditions, $\psi(n)=\psi(n+N_1+N_2)$ with a periodic $\theta$ field 
\begin{equation}
\begin{split}
&\theta(n)=\theta,\,n\in[kN,kN+N_1)\\
&\theta(n)=-\theta,\,n\in[kN+N_1,(k+1)N), k\in{\mathbb{Z}}
\end{split}
\end{equation}
Bloch’s theorem states that the general wave solution satisfies
\begin{equation}
\begin{split}
\psi(n,t)=e^{i\phi}\psi(n-N,t)
\end{split}
\end{equation}
We will use this Bloch phase $\phi$ to compensate the uniform $\alpha$ field. 
For now, we first find the bound state energy solution by matching the boundary conditions at $n=N_1$ and $n=N=N_1+N_2$ 
\begin{equation}
\begin{split}
\sin{E}=\pm\frac{(\mu^{-1}-\mu)\cos\theta\sqrt{\mu^{2N_1}+\mu^{2N_2}-2\mu^{N_1+N_2}\cos\phi}}{2\sqrt{(1-\mu^{2N_1})(1-\mu^{2N_2})}}
\end{split}
\end{equation}
For the symmetric ring with $N_1=N_2=M$, we simplify this exact result to 
\begin{equation}
\begin{split}
\sin{E}=\pm\frac{(\mu^{-1}-\mu)\cos\theta\sin{\frac{\phi}{2}}\mu^M}{1-\mu^{2M}}
\end{split}
\end{equation}

\begin{figure}[h]
\centering
\includegraphics[width=6cm]{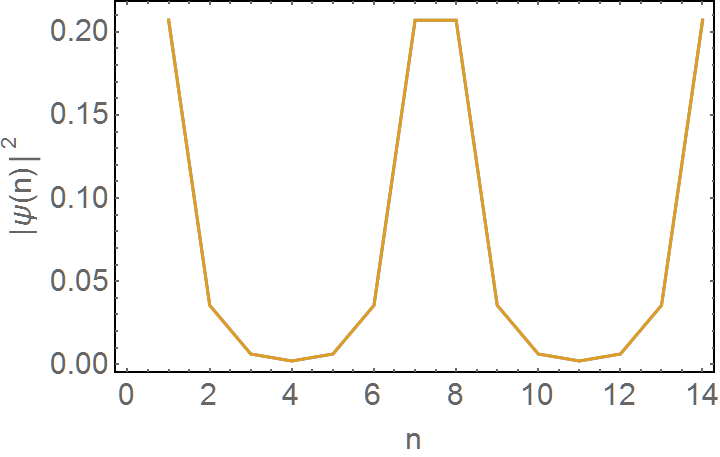}
\caption{The probability distribution $|\psi(n)|^2$ of the two normal bound states $E=\pm\epsilon_0$ , on the symmetric ring $N_1=N_2=7$. $\theta$ is set to be $0.25\pi$, and $\alpha=0.01\pi$. The two wave functions provide the same probability distribution, delocalized at the two topological boundaries $n=0$ (equivalently $n=N_1+N_2=14$), and $n=N_1=7$.}\label{F2}
\end{figure}

Next we add a constant uniform $\alpha$ field into the system. 
Since the uniform $\alpha$ field only add an extra phase $e^{i\alpha{n}}$ to the wave function without changing the eigen-energy on the infinite line, we can treat the ring as infinite line with periodical boundary conditions and consider the relation between this extra phase with the Bloch phase. 
Without the magnetic field, the ring system with two Majoranas corresponds to $\phi=0$. 
After adding $e^{i\alpha{n}}$, the continuity equation on the ring requires the flux quantization condition $2M\alpha+\phi=2m\pi,m\in\mathbb{Z}$ is satisfied. 
In order to match this boundary condition, we set the Bloch phase $\phi=2m\pi-2M\alpha,m\in\mathbb{Z}$. 
The Bloch phase will change after the addition of the $\alpha$ field in order to match the continuity condition of the wave function for the ring. 
We thus find that the ring with uniform constant $\alpha$ field corresponds to the periodic infinite line with initial Bloch phase $\phi=2m\pi-2M\alpha,m\in\mathbb{Z}$,
\begin{equation}
\begin{split}
\sin{E}=\pm\frac{(\mu^{-1}-\mu)\cos\theta\sin(M\alpha)\mu^M}{1-\mu^{2M}}
\end{split}
\end{equation}
For weak field $\alpha\ll{1}$, the bound state energy E is close to $0$ or $\pi$. Using the leading order approximation from dispersion relationship, we get
\begin{equation}
\begin{split}
&E=\pm\epsilon_0, \pi\pm\epsilon_1,\,\,\,\mu=\frac{\cos\theta}{1+\sin\theta}\\
&\epsilon_0=\epsilon_1=\pm{2}\mu^{M}\sin(N\alpha/2)\sin\theta,
\end{split}\label{E15}
\end{equation}
This approximation holds as long as $\alpha\ll{1}$. 
The corresponding eigenstates are the delocalized state around the two boundaries. 
For the weak perturbation, the delocalized states are approximately symmetric (Fig.\ref{F2}). 

\section{quantum oscillation on the ring}
When $\alpha$ field is absent, two Majoranas $E=0$ are localized at the two topological boundaries $n=0,M$. 
Let’s consider a system with a wave function $\psi(n,0)_{0}^{+}$ at $t=0$ given by one of the degenerate Majorana states, localized around $n=0$. The subscript 0 on  $\psi(n,0)_{0}^{+}$  indicates that initially  $\alpha =0$ and the $+$ sign refers to the Majorana state at $n=0$.
At time $t=0^+$, we add a non-zero weak $\alpha\ll{1}$ field. 
After adding $\alpha$ field, we express the Majorana wave function as the linear combination of the two normal bound states, $\psi(n,0)_{0}^{+}=c_{+}\psi^{+}_{{\alpha}}(n,0)+c_{-}\psi^{-}_{{\alpha}}(n,0)+R$, 
where $ \psi^{\pm}_{\alpha}$ are the two normal bound states $E=\pm\epsilon_0$ after adding ${\alpha}$ and the term $R=\sum_{i\neq{\pm}}{c_i}\psi^{i}_{\alpha}$ describes the projection to all those states that are not these two boundary modes. 
If we assume that $R$ is small, then we can describe the evolution of the system after the addition of the nonzero vector potential $\alpha$ by a new state with a two-level system composed only with the two ground states $ \psi^{\pm}_{\alpha}$. 
The error of this two-level approximation is due to the omission of $R$. 
We estimate the error by the inner product of $\psi_0^{+}(n,0)$ with other higher energy levels. 
For $\alpha=0.05\pi,\theta=0.25\pi$, and arbitrary $N_1=N_2=M>2$, we obtain $|R|^2<2\%$, indicating that the two-level approximation holds for a varieties of symmetric rings.
Note that the new wave function under a non-zero weak $\alpha\ll{1}$ field will oscillate as it is no longer a stationary state after the perturbation. 
The frequencies of the oscillation modes are equal to the energy gaps between different energy eigenstates of the system with $\alpha >0$. 
To compute this frequency in our two-level system approximation, we note that the only energy gap is $2E(\alpha)$. 
Thus, the two level approximation yields the period of the oscillation using Eq.\ref{E15} as  
\begin{equation}
\begin{split}
T=\frac{\pi}{{2}\mu^{M}\sin(N\alpha/2)\sin\theta},\,\,  \mu=\frac{\cos\theta}{1+\sin\theta}, 
\end{split}
\end{equation}

Consistent with our two-level approximation, we notice that the normal bound state wave function $ \psi^{\pm}_{\alpha}$ is approximately symmetric. Therefore we can absorb the phase of the new wave function by redefining the two normal bound states so that $c_{\pm}\approx{\frac{1}{\sqrt{2}}}$ to yield $\psi(n,0)_{0}^{+}\approx\frac{1}{\sqrt{2}}\psi^{+}_{{\alpha}}(n,0)+\frac{1}{\sqrt{2}}\psi^{-}_{{\alpha}}(n,0)$. 
The error of this approximation can be measured again by the mismatch $R'=\psi(n,0)_{0}^{+}-\frac{1}{\sqrt{2}}\psi^{+}_{{\alpha}}(n,0)+\frac{1}{\sqrt{2}}\psi^{-}_{{\alpha}}(n,0)$. 
For $\alpha=0.05\pi,\theta=0.25\pi$, and $N_1=N_2=M\leq{4}$, we obtain $|R|^2<5\%$. 
In general, this approximation holds better for a larger system.
The Majorana state $E=0$ around $n=M$ can similarly treated and we approximated it by 
$\psi(n,0)_{0}^{-}\approx\frac{1}{\sqrt{2}}\psi^{+}_{{\alpha}}(n,0)-\frac{1}{\sqrt{2}}\psi^{-}_{{\alpha}}(n,0)$, 
so that it is orthogonal to the Majorana state around $n=0$. 
If the system starts as a Majorana around $n=0$, the state will first evolve to the Majorana state around $n=M$ after half of the oscillation period. After a whole period, it will return to its original state. 
If we turn off the $\alpha$ field at $t=T/2$, then we obtain a stable Majorana around $n=M$.

\begin{figure}[h]
\centering
\includegraphics[width=6cm]{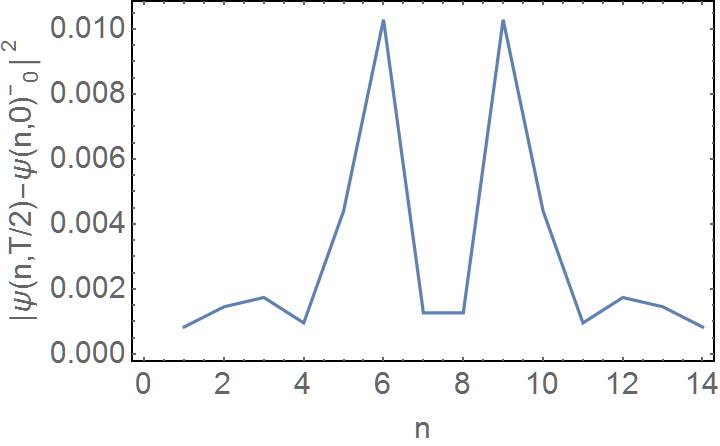}
\caption{The mismatch at every site between the normalized state after the oscillation of half period (64 steps)  and the true Majorana state. The simulation is on the symmetric ring $N_1=N_2=4$, with $\theta =0.25\pi$, and $\alpha=0.05\pi$.}\label{F3}
\end{figure}

Our numerical results show that for Hadamard coin with $\alpha=0.05\pi$, the loss of probability $|R'|^2$ is less than 5$\%$ for $M=4$. The oscillation period is $T\approx{128.1}$. The initial wave function is the Majorana state around $n=0$. 
After $t=64$ steps, we expect that the resulted state is the Majorana state around $n=M$. 
We show in Fig.\ref{F3} the mismatch $|\psi(n,T/2)-\psi(n,0)_{0}^{-}|^2$ of the resulted state and the true Majorana state at every site. The total mismatch between the resulted state $\psi(T/2,n)$ and the true Majorana state around $n=M$ is $\sum_n|\psi(n,T/2)-\psi(n,0)_{0}^{-}|^2\approx{10\%}$. 

We now can use the oscillation induced by the weak $\alpha$ field to transfer one Majorana state to another. The difference from the true Majorana state is only 10$\%$. This difference is smaller for a larger system, but the oscillation period is correspondingly longer. For example, for $M=5$ system shows less than $4\%$ difference, but the steps needed are doubled. 
We summarize the simulation results of the total mismatch and period T in Table. \ref{T1}. 
One can also tune the coin parameter $\theta$, so that the oscillation period is approximately an integer. This can also reduce the difference. 
\begin{table}[h]
\centering
\begin{tabular}{|c|c|c|c|}
\hline
M & $\alpha$   & T/2 & Total mismatch  \\ \hline
4 & $0.05\pi$  & 64  & $10.3\%$                                     \\ \hline
4 & $0.025\pi$ & 122 & $8.3\%$                                      \\ \hline
5 & $0.05\pi$  & 129 & $3.8\%$                                      \\ \hline
5 & $0.025\pi$ & 238 & $2.8\%$                                      \\ \hline
6 & $0.05\pi$  & 272 & $2.3\%$                                      \\ \hline
6 & $0.025\pi$ & 484 & $2.0\%$                                      \\ \hline
4 (10-step sequence) & $0.05\pi$ & 64 & $4.7\%$                                      \\ \hline
\end{tabular}
\caption{The total mismatch $\sum_n|\psi(n,T/2)-\psi(n,0)_{0}^{-}|^2$ between the evolved wave function and the true Majorana state, after operating $C(\pi/4,\alpha)$ for $T/2$. In general, the total mismatch is smaller for larger system with smaller $\alpha$.}\label{T1}
\end{table}

Now we obtain two operations to manipulate the two Majorana fermion. 
We denote the Majorana at $n=0$ as $|0\rangle$, and Majorana at $n=M$ as $|1\rangle$. 
If the operator $C_0=C(\theta,\alpha=0)$ is applied, the Majoranas will not change, so $C_0|0\rangle=|0\rangle, \,C_0|1\rangle=|1\rangle$. 
The second operation $C_1$ is to apply $C(\theta,\alpha\neq0)$ for $T/2$ steps. 
Then the two Majorana switches, i.e. $C_1|0\rangle=|1\rangle, \,C_1|1\rangle=e^{i\phi'}|0\rangle$. Here $\phi'$ is an extra phase.
A sequence of these two operations, applied by turning off and on the $\alpha$ field, actually provide decent result with low total mismatch. For example, for a ring with $M=4$ and $\theta=0.25\pi, \, \alpha=0.05\pi$, the sequence $C_1C_0C_1C_1C_0C_1C_1C_0C_1C_1$ (start from left to right), transforms $|0\rangle$ to $|1\rangle$ with a total mismatch $\sum_n|\psi(n,T/2)-\psi(n,0)_{0}^{-}|^2=4.7\%$. We see that this combination of $C_0$ and $C_1$ provide a valid way to manipulate the two Majorana states for multiple times and the error is smaller than 10$\%$ (refer to Table. \ref{T1}.)

\section{Discussion}
We exploit the topological boundary modes found for quantum wire and consider their behavior for a finite ring. We find for the symmetric ring two quasi-particles that can be identified as the Majorana states\citep{Hotat, Kitaev,Kitagawa}. We also discuss the approximate solution for these Majorana states for asymmetric ring. For symmetric ring of size 2M, we have the exact solution for the Majorana states. Our solution simplifies the experimental setup for small system such as in cold atom or photon system and provide a way to reduce decoherence. We also consider the effect of the states after turning on a uniform constant $\alpha$ field. We find that the initial Majorana state can be decomposed into normal bound states and starts to oscillate with a frequency equal to the energy gap $2E(\alpha)$ in our two-level approximation which is valid when $\alpha$ is sufficiently less than $1$. This field induced oscillation provides a possible way to transfer one Majorana state to another, with a characteristic time $T/2=\pi/E$. The numerical result on small system demonstrates the validity of this evolution, within reasonable steps. The information loss is around 10$\%$ for $M=4$, and can be further suppressed for larger system and suitably chosen coin parameter $\theta$. 
We expect the one-dimensional quantum walk on finite ring can be realized in a trapped-ion chain with an Majorana qubit encoded \citep{Mezzacapo}. 
This qubit is topologically protected against major sources of decoherence, thereby providing an efficient quantum memory. 
The oscillation of these symmetry protected bound states provides a new phenomenon that may be useful in quantum computation.

\begin{acknowledgments}
We acknowledge discussion with Lam Hotat and Wang Ning.
\end{acknowledgments}
\bibliography{bound}

\end{document}